\documentclass[conference]{IEEEtran}
\usepackage[utf8]{inputenc}
\ifCLASSINFOpdf
  \usepackage[pdftex]{graphicx}
  \graphicspath{{./}}
  \DeclareGraphicsExtensions{.pdf}
\else
\fi
\usepackage[cmex10]{amsmath}
\usepackage[caption=false,font=footnotesize]{subfig}
\usepackage{amssymb}
\usepackage{todonotes}
\usepackage{acronym}

\acrodef{IP}{Internet Protocol}
\acrodef{DNS}{Domain Name System}
\acrodef{TLS}{Transport Layer Security}
\acrodef{CPRNG}{Cryptographic Pseudo-Random Number Generator}
\acrodef{GPS}{Global Positioning System}
\acrodef{geocast}{Geographic Multicast}
\acrodef{mobicast}{Just-in-time Multicast}
\acrodef{STM}{Spatiotemporal Multicast}
\acrodef{GTM}{Geo-temporal Multicast}
\acrodef{CSTM}{Cluster-based Spatiotemporal Multicast}
\acrodef{SMS}{Short Message Service}
\acrodef{LTE}{Long-Term Evolution}
\acrodef{S}{Sender}
\acrodef{UE}{User Equipment}
\acrodefplural{UE}[UEs]{User Equipments}
\acrodef{eNB}{evolved NodeB}
\acrodef{HSS}{Home Subscriber Server}
\acrodef{MME}{Mobility Management Entity}
\acrodef{UE}{User Equipment}
\acrodef{PGW}{Packet Data Network Gateway}
\acrodef{IMSI}{International Mobile Subscriber Identity}
\acrodef{TMSI}{Temporary Mobile Subscriber Identity}
\acrodef{NAT}{Network Address Translation}
\acrodef{RP}{Rendezvous Point}
\acrodefplural{RPs}[RPs]{Rendezvous Points}
\acrodef{DoS}{Denial-of-Service}
\acrodef{TPS}{Token Planning Server}
\acrodef{st-r}[$st$-region]{spatiotemporal region}
\acrodef{st-d}[$st$-datagram]{spatiotemporal datagram}

\makeatletter
\definecolor{micksgreen}{rgb}{0,0.5,0}
\definecolor{micksgreenlight}{rgb}{0,0.6,0}
\newcommand{\m}[1]{\color{micksgreen}#1 \color{black}}
\newcommand*{\cM}{\@ifstar\cM@star\cM@nostar}
\newcommand*{\cM@star}[1]{\todo[size=\tiny,color=micksgreenlight]{\sf #1}}
\newcommand*{\cM@nostar}[1]{\todo[inline,size=\scriptsize,color=micksgreenlight]{\sf #1}}
\definecolor{sander}{rgb}{1,0.4,0}
\definecolor{sanderlight}{rgb}{1,0.5,0}

\newcommand*{\cS}{\@ifstar\cS@star\cS@nostar}
\newcommand*{\cS@star}[1]{\todo[size=\tiny,color=sanderlight]{\sf #1}}
\newcommand*{\cS@nostar}[1]{\todo[inline,size=\scriptsize,color=sanderlight]{\sf #1}}
\makeatother

\newcommand{\itembf}[1]{\item\textbf{#1:} }

\usepackage[absolute]{textpos}
\usepackage{calc}
\newcommand{\copyrightnotice}{
    \begin{textblock}{0.78}(0.11,0.03)
        \noindent
        \footnotesize
        \textcolor{gray}{
        {\bf This paper is a preprint (IEEE "accepted" status).} 
        ~\copyright~2013 IEEE. 
        Personal use of this material is permitted.
        Permission from IEEE must be obtained for all other uses, in any current or future media, including reprinting/republishing this material for advertising or promotional purposes, creating new collective works, for resale or redistribution to servers or lists, or reuse of any copyrighted component of this work in other works.
        }
    \end{textblock}
}

\begin{document}
%
\title{Geocast into the Past: Towards a Privacy-Preserving Spatiotemporal Multicast for Cellular Networks}



\author{
\IEEEauthorblockN{Sander Wozniak, Michael Rossberg, Franz Girlich, Guenter Schaefer}
\IEEEauthorblockA{Ilmenau University of Technology\\
\{sander.wozniak, michael.rossberg, franz.girlich, guenter.schaefer\}[at]tu-ilmenau.de}
}


%


\maketitle

\copyrightnotice

\begin{abstract}
This article introduces the novel concept of \acf{STM}, which is the issue of sending a message to mobile devices that have been residing at a specific area during a certain time span in the past.
A wide variety of applications can be envisioned for this concept, including crime investigation, disease control, and social applications.
An important aspect of these applications is the need to protect the privacy of its users.
In this article, we present an extensive overview of applications and objectives to be fulfilled by an \ac{STM} service.
Furthermore, we propose a first \ac{CSTM} approach and provide a detailed discussion of its privacy features.
Finally, we evaluate the performance of our scheme in a large-scale simulation setup.
\end{abstract}


%
\IEEEpeerreviewmaketitle

\section{Introduction}

\begin{figure}[!t]
\centering
\includegraphics{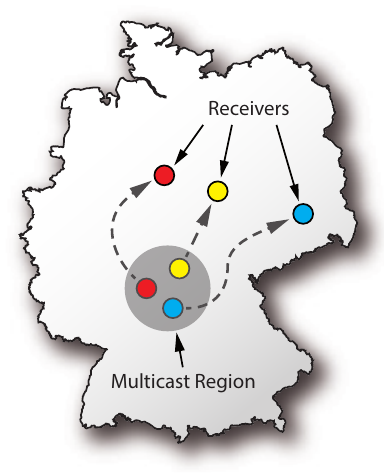}
\caption{Example for the concept of \acf{STM}.}
\label{fig:concept}
\end{figure}

In this work, we introduce the novel concept of \acf{STM}, i.e., the issue of sending a message to all cell phones that have been residing at a specific area during a certain time period in the past (see Fig.\;\ref{fig:concept}).
This concept can be considered as an extension of \ac{geocast}~\cite{maihofer2004survey}, introducing the temporal aspect of sending a \ac{geocast} message ``into a certain point in the past''.

We envision a wide variety of applications for the sketched service.
For example, it may be used to discover witnesses of a crime by sending a message to mobile phones that have been near the crime scene at the time the crime has presumably been committed.
In addition, an \ac{STM} service could help in controlling the outbreak of an infectious or contagious disease, as carriers of a pathogen are often unaware of their contamination.
With incubation times delaying the appearance of first symptoms several hours, days, or weeks, it is hard to know who has been infected.
While an early detection of carriers is vital to successfully treating a disease, it is usually impossible to quickly test large parts of the population for infection.
Assuming the area of the outbreak or the path of one single carrier is roughly known (e.g.\m{,} by evaluating the past movement of a victim), an \ac{STM} service could advise other potential victims to seek medical attention.
Finally, an \ac{STM} service may be used for mobile social networking applications, e.g., in order to exchange information or to get in contact with people encountered in the past.
These applications could range from exchanging pictures with fellow visitors after a festival to dating services enabling people to meet again after a random encounter where no contact information has been exchanged.

An important aspect of these applications is the privacy of the users receiving \ac{STM} messages.
For example, a witness may not want to reveal his presence to the crime scene to someone else, e.g., out of fear of retaliation by the perpetrator.
A person potentially infected with a disease may want to be alerted without having revealed his identity to someone else than his doctor.
Finally, for social applications, users may not want to be identified by the sender or other receivers, being free to ignore or react to the content of the message.

A na\"{\i}ve approach for realizing a privacy-preserving \ac{STM} is to let each mobile record its own movement.
Then, a sender can deliver a message by broadcasting it to all users.
A mobile only shows the received message to its user, when its recorded path intersects with the destination time and place.
It is easy to see that, in this scenario, a sender has to make the information in the message available to everyone.
However, a sender may desire to keep the information confidential between him and the users in the destination region.
In this case, only users that have actually been present in the area at the time should receive the message.
Therefore, a privacy-preserving \ac{STM} approach should deliver messages only to users having resided in the spatiotemporal destination region.

There are several research areas related to the concept of \ac{STM}.
The issue of \acf{geocast} has been studied extensively since the past decade \cite{maihofer2004survey}.
Nevertheless, it does not consider the temporal aspect of sending a message to nodes that have been residing in an area at a point in the past.
In contrast, the concept of \ac{mobicast} does consider the spatiotemporal aspect \cite{huang2003mobicast}.
However, its goal is to deliver a message to a geographic area residing in the future path of a node for just-in-time delivery.
Therefore, it only considers the temporal aspect of sending a message to an area at a point in the future, not to the past.
Furthermore, there exists a wide variety of research on the issue of spatiotemporal querying and indexing structures \cite{pelanis2006indexing,meka2005dist}.
These approaches, however, focus on the issue of efficiently querying regions in space and time and do not incorporate the aspects of privacy or message delivery (i.e. multicast).
Finally, while there has been some work on privacy-preserving spatiotemporal querying \cite{ku2009privacy}, it does not consider the issue of privacy-preserving multicast.

Within this article we make the following contributions:
\begin{itemize}
    \item the concept of spatiotemporal multicast is introduced,
    \item extensive privacy objectives are derived,
    \item we present a first method to realize a privacy-preserving spatiotemporal multicast service for cellular networks,
    \item and evaluate it by a comprehensive simulative evaluation.
\end{itemize}

The rest of the article is organized as follows:
In section~\ref{sec:objectives} we present objectives to be fulfilled by a \ac{STM} service.
Section~\ref{sec:approach} describes our \ac{CSTM} scheme, while a discussion of its privacy features is given in section~\ref{sec:discussion}.
Finally, we present the results of the performance evaluation of our approach in section~\ref{sec:evaluation} and conclude with an outlook in section~\ref{sec:conclusion}.

Please note that we use the term \ac{st-r} to refer to a geographic area during a certain time period.
Furthermore, we use the term \ac{st-d} when referring to a message that is to be delivered to users in such a \ac{st-r}.
For components of the cellular network, we use the terminology of \ac{LTE}.

\section{Design Objectives}
\label{sec:objectives}
\subsection{Functional Objectives}

\begin{itemize}
    \itembf{Long-term support} A sender should be able to send a message to users having resided in the area a long time ago.
        Given the above mentioned applications, this time can range from several hours or days to even weeks.
    \itembf{Delivery ratio} Users present at the addressed \ac{st-r} should receive a \ac{st-d} with high probability.
    \itembf{Precision} Senders should be able to address a \ac{st-r} up to several minutes within specific radio cells.    
\end{itemize}

\subsection{Non-functional Objectives}

\begin{itemize}
    \itembf{Delivery speed} The delivery delay (the time between sending and receiving) should be low.
    
    \itembf{Efficiency} The service should be efficient in terms of computation, memory, and communication overhead.
        The latter includes that only users having actually been in the destined \ac{st-r} should receive the \ac{st-d}.
    \itembf{Scalability} The objectives mentioned above should not be significantly degraded by an increasing number of users and senders (i.e. number of \acp{st-d}).
    \itembf{Robustness} The service should deliver \acp{st-d} with high probability despite failures of the infrastructure.
\end{itemize}

\subsection{Privacy and Security Objectives}

For user privacy, the following aspects must be considered:
\begin{itemize}
    \itembf{Anonymity}
        Attackers must not infer the identity of users receiving a \ac{st-d}.
        Otherwise, they could unwillingly be associated with events addressed in it.
    
    \itembf{Location privacy} 
        Attackers must not infer the past, present, or future locations of users up to a defined accuracy.
        This is important as knowing the location of users during the night time can reveal their home addresses, for example.
    
    \itembf{Co-location privacy}
        Attackers must not be able to decide whether two users have been residing in the same radio cell at the same time.

    \itembf{Absence privacy} 
        Attackers must not determine the absence of a user from a \ac{st-r} (e.g. by automatically testing whether he receives a message addressed to this region).
        This knowledge can be harmful if a user was not at a specific \ac{st-r} although he was supposed to be.

    \itembf{Data confidentiality} Only users having resided in the destined \ac{st-r} should be able to read and detect the availability of a message for this region. This is required by some applications, as detecting an a message can already compromise the confidentiality of the information (e.g. in case of a rare event like the outbreak of a disease).

\end{itemize}

\section{Approach}
\label{sec:approach}

In this article, we propose a \acf{CSTM} approach which uses \emph{\acp{RP}} to deliver \acp{st-d}.
These \acp{RP} act like mailboxes where senders can deposit \acp{st-d}, allowing users to retrieve them later by polling the \acp{RP} for the \acp{st-r} they have been residing in.
Thus, the message can also be retrieved if the \emph{\ac{UE}} has not been available at the time the message has been sent.
It also allows \acp{RP} to bundle multiple \acp{st-d} in one packet when responding to polls.

The confidentiality of a message between a sender and its intended receivers requires a key exchange.
Since a key can only be exchanged proactively for future messages, we use the base stations, referred to as \emph{\acp{eNB}}, to perform a key exchange.
Therefore, the base stations generate and broadcast tokens~$\tau$ containing symmetric keys~$K$ at regular time intervals $t$.
Base stations generate these keys from an initial random seed~$s$ using a \ac{CPRNG}.
In order to generate these seeds, a trusted entity is required.
This entity, referred to as \ac{TPS}, is able to recover keys using its knowledge of seeds and time slots at which keys are generated by \acp{eNB}.
Generating the keys on the \acp{eNB} has the advantage that online interaction with the \ac{TPS} is only necessary when sending \acp{st-d}.
Furthermore, adversaries are usually not able to compromise \acp{eNB} without being noticed.

The \acp{UE} use the keys contained in the tokens to determine which \acp{RP} to poll and which \ac{st-r} identifiers to supply for message lookup.
\acp{UE} obtain this \ac{st-r} identifier $r_K=h(K)$ by applying a cryptographic hash function $h(x)$ to the key $K$.
This identifier $r_K$ allows \acp{RP} to perform a lookup for \acp{st-d} while preventing adversaries from obtaining $K$ using a compromised \ac{RP}.
In order to determine which \acp{RP} to poll with $r_K$, \acp{UE} apply $h(x)$ to $r_K$.
From this \ac{RP} identifier $rp_K = h(r_K) = h(h(K))$, \acp{UE} obtain the name of the \ac{RP} to be polled.
\acp{UE} compute the name by appending the number $rp_K \bmod N$ to a known prefix, where $N$ is the known number of \acp{RP}.
Given this name, the IP address is resolved using the \ac{DNS}.
This approach distributes the load evenly among the \acp{RP} while preventing attackers from inferring the \acp{st-r} being stored on a \ac{RP}. 

\begin{figure}[!t]
\centering
\includegraphics{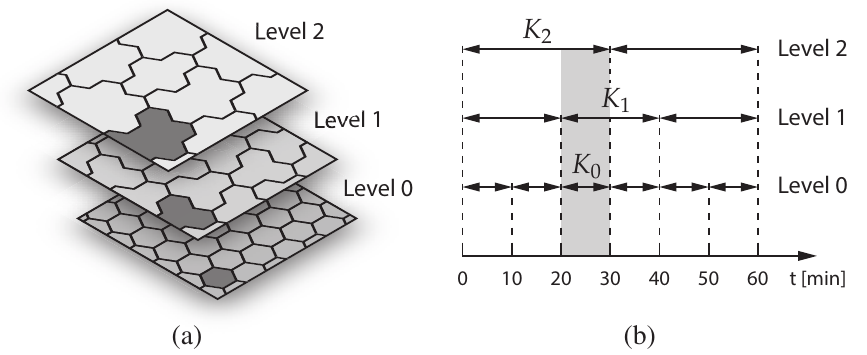}
\caption{Examples of token hierarchies in (a)~space and (b)~time.
Each spatial cluster and time interval shown here uses a different key.
The darker areas indicate which keys are part of a token~$\tau_{c,t}$.}
\label{fig:aggregation}
\end{figure}

A potential issue of the \ac{CSTM} approach is the polling overhead.
When polling \acp{RP}, \acp{UE} check for messages for all keys contained in the tokens that they have received so far (using \ac{st-r} and \ac{RP} identifiers as described above).
Hence, the polling overhead is directly related to the number of keys a \ac{UE} has received.
In order to reduce the number of polling messages between \acp{UE} and \acp{RP}, we introduce the concept of a \emph{token hierarchy}.
A token hierarchy is used to aggregate tokens over space and time (see Fig.\;\ref{fig:aggregation}), providing a tradeoff between delivery accuracy and polling overhead (evaluated in section~\ref{sec:evaluation}).
This is achieved by distributing tokens that consist of several different symmetric keys (one for each level $l$ of all $\lambda$ levels of the hierarchy).
The token hierarchy $L = \left\{ (l, t^s_l, t^v_l) \mid 0 \leq l < \lambda \right\}$ defines the time slot sizes $t^s_l$ of a level and the continuous validity periods $t^v_l = [a_l,b_l)$ relative to the time a token has been received by a \ac{UE}.
Hence, $a_0 = 0$ and $\forall l, 0 < l < \lambda: b_{l-1} = a_l$.
When polling a \ac{RP}, a \ac{UE} uses the key based on the time that has passed since the reception of the token.
Each level of the hierarchy defines the validity period $t^v$ during which the key of this level is to be used.
For example, assuming $L = \left\{ (0, 10, [0,30)), (1, 20, [30,60)), (2, 30, [60,90)) \right\}$ (times in minutes), during the first 30~minutes after having received the token, a \ac{UE} uses level~0 when polling for messages.
Between 30~and~60~minutes after the reception, the key of level~1 is used and between 60~and~90~minutes, the \ac{UE} uses the key of level~2.
After that, the token is considered stale and deleted by the \ac{UE}.
Since with increasing levels, the granularity of the addressed \ac{st-r} decreases according to Fig.\;\ref{fig:aggregation}, the number of keys to be polled by a \ac{UE} should be decreased as well.
However, in this case, the delivery accuracy is expected to degrade.
Therefore, users not having resided in the region may receive messages not intended for them.
While decreasing the delivery accuracy is not desired regarding the confidentiality between sender and receivers, it improves the privacy of users by obfuscating their movement trails according to the concept of $k$-anonymity.

\begin{figure}[!t]
\centering
\includegraphics{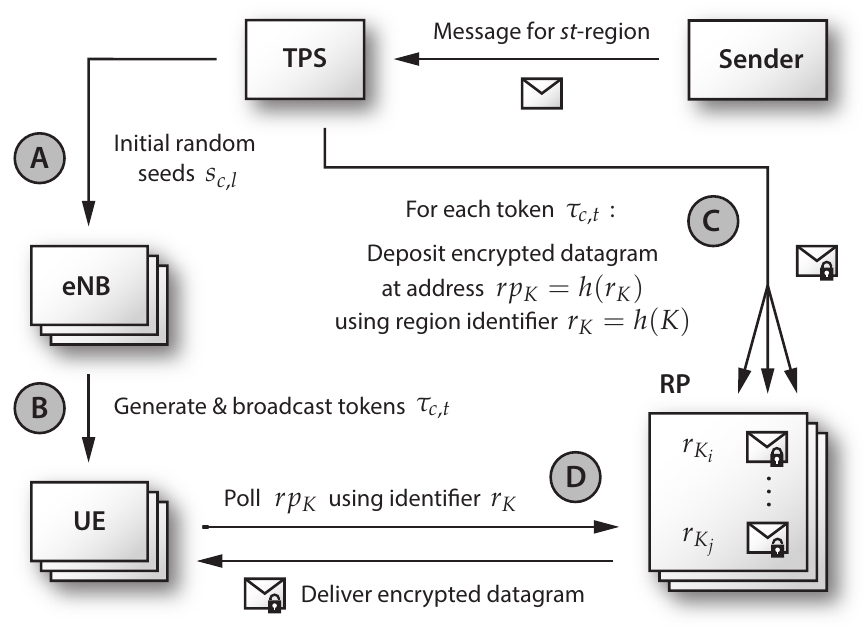}
\caption{Overview of the phases of the \ac{CSTM} approach: (A) token planning, (B) token distribution, (C) message deposition, and (D) message delivery.
Messages exchanged between entities are protected by \acs{TLS}.}
\label{fig:overview}
\end{figure}

We now provide a detailed overview of the four phases of the \ac{CSTM} approach in the following sections (see Fig.\;\ref{fig:overview}).

\subsection{Initialization / Token Planning}

In the first phase, the \ac{TPS} uses its knowledge of the layout of the cellular network to generate initial random seeds $s_{c,l}$ for each cell $c \in C$ and level $l$ of the token hierarchy.
Here, $C$ is the set of all radio cells. 
Base stations obtain these initial random seeds for key generation (e.g., given 3~levels, a base station will receive 3~initial seeds).
Base stations of cells grouped in a spatial cluster at a level obtain the same seed for this level (see Fig.\;\ref{fig:aggregation}).
Finally, the TPS sends the initial random seeds to the \acp{eNB} using a cryptographic protocol providing confidentiality, data integrity, and entity authentication, for example, using the \ac{TLS} protocol.

\subsection{Token Generation and Distribution}

As mentioned above, establishing message confidentiality requires a proactive key exchange.
Hence, tokens~$\tau_{c,t}$ are distributed to all \acp{UE} in the cell~$c \in C$ during each time slot~$t$.
In cell $c$ at the beginning of time slot $t$, an \ac{eNB} creates the token $\tau_{c,t} = \left\{ (t^v_l,K_{c,t,l}) \mid 0 \leq l < \lambda \right\}$ by generating the symmetric keys $K_{c,t,l}$ using the \acl{CPRNG} of level $l$ (\acs{CPRNG}$_l$) initialized with random seed $s_{c,l}$.
How often a new key is generated for a level $l$ depends on the time slot $t^s_l$ used at this level.
E.g., in Fig.\;\ref{fig:aggregation}b, an \ac{eNB} would generate a new key every 10~minutes at level~0, every 20~minutes at level~1, and every 30~minutes at level~2.
In this example, an \ac{eNB} would send the token $\tau_{c,t} = \left\{ (t^v_0,K_0), (t^v_1,K_1), (t^v_2,K_2) \right\}$ to all \acp{UE} residing in cell $c$ at some time during time slot $t = [20,30)$.

\subsection{Message Deposition}

When a legitimate sender wants to send a message to a specific \ac{st-r}, he sends the message to the \ac{TPS}, which encrypts and distributes it to the \acp{RP}.
Here, the connections between sender and \ac{TPS}, as well as between \acp{RP} and \ac{TPS} must be secured by a cryptographic protocol like \ac{TLS}.
The request of the sender contains a description of the intended \ac{st-r}, for example a rectangular area and a time interval.
If the sender can be authenticated and is authorized to send a \ac{st-d}, the \ac{TPS} looks up the initial random seeds for the cells spatially overlapping with the destination region.
Then, the \ac{TPS} uses these seeds and the known time spans $t^s_l$ at which keys are generated by \acp{eNB} to calculate the necessary tokens.

Since a token~$\tau_{c,t}$ has different keys $K_{c,t,l}$, the \ac{TPS} distributes the \ac{st-d} based on the level that is currently valid according to the validity periods $t^v$ of the levels.
It only uses keys that are currently valid and encrypts the message with each key $K$.
For example, given token hierarchy $L = \left\{ (0, 10, [0,30)), (1, 20, [30,60)), (2, 30, [60,90)) \right\}$ (times in minutes) and a token that has been announced 40~minutes ago, the \ac{TPS} will encrypt and deposit the \ac{st-d} using the key of level~1.
According to the procedure described above, the \ac{TPS} obtains the \ac{RP} identifier $rp_K = h(h(K))$ and the \ac{st-r} identifier $r_K = h(K)$ for each key $K$.
Again, the hash function is necessary to enable message lookup without revealing the key
The \ac{TPS} obtains the addresses of the \acp{RP} by resolving $rp_{K} \bmod N$ via \ac{DNS}, where $N$ is the known number of \acp{RP}.
Then, for each key $K$, the \ac{TPS} encrypts and sends one \ac{st-d} as well as $r_{K}$ to the respective \ac{RP}.
Finally, the \acp{RP} store the datagrams for lookup using $r_{K}$.

\subsection{Message Delivery}

\acp{UE} store the tokens received from the \acp{eNB} during their visit in the respective cells.
This trail of tokens is then used to check for new messages at regular time intervals.
When checking for messages, for each token~$\tau_{c,t}$, a \ac{UE} chooses the key based on the level that is currently valid according to the validity periods $t^v$.
It also obtains the \ac{st-r} identifier $r_{K} = h(K)$ from the key and the address of the \ac{RP} $rp_K = h(r_K)$.
Then, the \ac{UE} sends a polling message containing region identifier~$r_K$ to all \acp{RP} $rp_K$ (one polling message for each keys $K$).
\acp{RP} receiving a polling message perform a lookup for the given region identifier~$r_K$ and deliver all \acp{st-d} that are available.
Here, the connections between \acp{UE} and \acp{RP} are also protected by a cryptographic protocol like \ac{TLS}.
Finally, \acp{UE} decrypt the received \acp{st-d} with the respective symmetric key $K$.

\section{Privacy Discussion}
\label{sec:discussion}

As stated in section~\ref{sec:objectives}, we assume that potential attackers have one or more of the following goals:
to infer the identity of users receiving a \ac{st-d}, their locations, co-location, absence, or a plaintext of a \ac{st-d} for a region he has not been residing in.
Attackers may observe the communication between entities, employ \acp{UE}, and send \acp{st-d} in order to achieve these goals.
More sophisticated attackers might even compromise \acp{RP} and \acp{eNB}.
However, attackers may not compromise the \ac{TPS} and obtain access to the initial random seeds.
We consider this a legitimate assumption since it is easier to control access to one or only few \acp{TPS}, whereas a large number of \acp{RP} may be needed for scalability reasons.
Furthermore, we assume that attackers are not able to compromise the \ac{PGW}, \ac{HSS}, or \ac{MME} of the cellular operators.
Otherwise, an adversary would already be able to track the movement of all subscribed users and obtain their identity.
Given the above mentioned abilities of adversaries, we consider the following four attacks of increasing power.

\subsection{Observation Attack}
In the first potential attack, adversaries observe the communication between \ac{eNB} and \acp{UE}, \acp{UE} and \acp{RP}, \ac{TPS} and \acp{RP}, as well as sender and \ac{TPS}.
Due to the employed \ac{TLS} protocol, attackers may only violate the objectives of location, co-location, and absence privacy of the users within the cell he is currently residing in.
This, however, is not due to the \ac{STM} service, but the possibility to directly observe users in his surroundings.
Furthermore, attackers may only infer the identify of users from their location, as inferring the identity from IP~address, \ac{IMSI}, or \ac{TMSI} requires additional knowledge from the cellular operator.
Finally, due to his presence in the cell at the specific time, an adversary becomes a legitimate receiver.
Hence, obtaining the symmetric key $K$ does not violate the objective of confidentiality.

\subsection{Probing Attack}
In the second potential attack, adversaries observe the communication between \ac{TPS} and \acp{RP}, as well as between \acp{UE} and \acp{RP}.
By sending a \ac{st-d} to a specific region, he tries to infer the \ac{RP} that is responsible for the given \ac{st-r}.
Then, he tries to identify users having resided at this region by observing which \acp{UE} poll the respective \ac{RP}.
Here, an attacker must be able to recognize his message among other messages being exchanged between \ac{TPS} and \acp{RP}.
This, however, is only possible if he is the only one sending a \ac{st-d} as the employed \ac{TLS} protocol does not allow him to decrypt these messages.
Nevertheless, even if an adversary can obtain the corresponding \ac{RP} identifier $rp_K$, he is not able to obtain the region identifier $r_K$ due to the preimage resistance of $h(x)$.
Accordingly, he cannot obtain $K$ without having resided in the intended \ac{st-r}.
Given these constraints, we now provide a detailed discussion of each privacy and security objective:

\subsubsection{Anonymity}
In this attack, adversaries are not able to infer the identity of users from their IP address as this requires access to additional knowledge from the cellular operator.

\subsubsection{Location privacy}
Attackers are also not able to obtain the locations of receivers as they may only learn about the \acp{RP} that are responsible for certain \acp{st-r}.
However, as \acp{RP} are responsible for a large number of different \acp{st-r} in an unpredictable manner according to $rp_K = h(r_K)$, they can not infer the \acp{st-r} addressed by \acp{UE} polling a \ac{RP}.

\subsubsection{Co-location privacy}
According to the objective of location privacy, attackers can also not learn whether two \acp{UE} polling the same \ac{RP} have been residing in the same \ac{st-r}.

\subsubsection{Absence privacy}
Adversaries can only learn about the absence of users from a \ac{st-r} if the corresponding \ac{RP} is not being polled by a \ac{UE}.
This, however, is not the case as \acp{RP} are responsible for a large number of different \acp{st-r}.

\subsubsection{Confidentiality}
Attackers are not able to obtain $K$ unless they have resided in the addressed \ac{st-r}.

\subsection{Compromising \acp{RP}}
More sophisticated adversaries may also compromise one or more \acp{RP}.
This enables attackers to gain access to the region identifiers $r_K$ mapped to this \ac{RP}.
However, they are still not able to obtain the original \acp{st-r} of these identifiers due to the preimage resistance of $h(x)$.
In order to obtain the mapping $r_K \rightarrow $~\ac{st-r}, adversaries have to rely on probing messages as described above.
Therefore, they have to guess a \ac{st-r} which is mapped to the compromised \ac{RP}.
Please note that guessing a \ac{st-r} requires the adversaries to actually send probing messages with different regions via the \ac{TPS}.
This is due to the fact that they can not guess keys $K$ for regions resulting in an identifier $r_K = h(K)$ located on the compromised \ac{RP}. 
Furthermore, in their probing messages, attackers have to address a single cell $c$ during one time slot $t^s$ to obtain a unique mapping of $r_K$ to \ac{st-r}.
This results in a potentially large number of probing messages.
Since authenticated senders cannot perform a Sybil attack, a \ac{TPS} can detect and filter probing messages (such a mechanism is also needed to prevent spamming and \ac{DoS} attacks).
Moreover, due to the token hierarchy, messages are delivered to different \acp{RP} using different \ac{st-r} identifiers~$r_K$ based on the current level~$l$.
Since $l$ depends on the time that has passed since the destined \ac{st-r}, attackers only have a limited time for probing \acp{st-r} before a higher level and thus different \acp{RP} are responsible.
Given these constraints, we now discuss the given objectives individually:

\subsubsection{Anonymity}
According to the previous attack, adversaries are not able to infer the identity of users from their IP address as this requires knowledge from the cellular operator.

\subsubsection{Location privacy}
Attackers may be able to partially violate the location privacy of users.
This, however, requires that they are able to obtain the mapping of $r_K$ to \ac{st-r}.
Furthermore, while adversaries may infer a specific \ac{st-r} where several anonymous users have been residing, they are not able to track their movements over several time slots and cells unless they control all \acp{RP}.
This is due to the fact that different \acp{st-r} are mapped to \acp{RP} using $h(x)$.

\subsubsection{Co-location privacy}
With this attack, the co-location privacy of users may be violated by detecting whether two users poll using the same $r_K$.
Nevertheless, in order to find out if two known users have been neighbors at some time, attackers have to be able to specifically compromise the \ac{RP} responsible for the assumed \ac{st-r} of the meeting.
Furthermore, they have to know the current IP addresses of both \acp{UE} to detect whether these \acp{UE} poll using the same $r_K$.

\subsubsection{Absence privacy}
Attackers may violate the absence privacy of users with this attack by detecting whether a user does not poll a certain $r_K$.
However, in order to find out if a known user has been absent from a specific \ac{st-r}, an attacker has to know the current IP address of the user and the $r_K$ of the relevant region.
Moreover, due to the token hierarchy, attackers have to be able to specifically compromise the \acp{RP} responsible at the current level.

\subsubsection{Confidentiality}
Adversaries are not able to obtain $K$ from $r_K = h(K)$ due to the preimage resistance of $h(x)$.

\subsection{Compromising \acp{eNB}}
Assuming that attackers are able to compromise \acp{eNB}, they may violate all given objectives for users in this region.
Although it may be rather unlikely for attackers to compromise \acp{eNB} without being noticed, the approach is still able to provide graceful degradation.
Hence, the violation is limited to the cell, affecting only \acp{st-r} after the compromise.

\section{Performance Evaluation}
\label{sec:evaluation}

We evaluated the efficiency of our scheme in terms of the polling overhead and the delivery accuracy.
Therefore, we implemented \ac{CSTM} using OMNeT++ \cite{varga2001omnet++} and INET.
We built a simplified architecture of a cellular network consisting of \acp{UE}, \acp{eNB} and a \ac{PGW} connected to a router and the \acp{RP}.

We modelled the mobility of users using SUMO \cite{behrisch2011sumo} and the TAPAS Cologne scenario (between 6~and~8\,am) \cite{uppoor2011large}.
On one hand, we chose this vehicular scenario, because it provides realistic mobility for a large-scale setting with over 100\,000 vehicles.
On the other hand, we assume that \acp{UE} travelling at a speed of about 50\,km/h present a high-load situation as \acp{UE} receive a lot of tokens.
We extended SUMO in order to incorporate the locations of \acp{eNB} and used a Voronoi diagram of \ac{eNB} coordinates to model their coverage.
In order to obtain a trace of cell switches, we calculated the intersection between the positions of vehicles and the Voronoi cells.
We used this simple approach, as we are not interested in the points where \acp{UE} switch cells, but in the number of tokens that they receive.

The \ac{eNB} coordinates were obtained from a website collecting such locations for Germany (http://www.senderliste.de/).
We converted the given street addresses to geo-coordinates using Nominatim (http://nominatim.openstreetmap.org) and adjusted them to the x,y-coordinates of the Cologne scenario.

In our simulation, the 604~\acp{eNB} announced tokens every 3~minutes or delivered them when \acp{UE} entered their cell, whereas \acp{UE} polled \acp{RP} every 10~minutes.
We used a token hierarchy with $\lambda=4$ levels $L = \left\{ (0, 3, [0,15)), (1, 6, [15,30)), (2, 9, [30,60), (3, 12, [60,120)) \right\}$ (times in minutes).
Hence, for temporal aggregation, the time step was increased by 3~minutes with each level $l$ ($0 \leq l < 4$).
For spatial aggregation, we used random clustering to distribute the same initial seed to the corresponding \acp{eNB}, grouping them into random $k$-clusters at level~$l$ ($k = l + 1$).

For each of the 32~repetitions, we let a sender send 100~\acp{st-d} to a randomly chosen \ac{st-r} (uniform distribution).
Its rectangular shape was randomly chosen anywhere in the city.
The addressed time span was $[a,b]$, where $a$ was randomly chosen from the interval $[0,5]$ and $b$ from $[5,10]$.
Hence, the duration of the \ac{st-r} was between 5~and~10~minutes.
The sender waited a random time between 10~and~60~minutes after $b$ before sending the \ac{st-d}.

We evaluated the following scenarios: (N) no token hierarchy, (S) only spatial aggregation, (T) only temporal aggregation, and (ST) both spatial and temporal aggregation.

\begin{figure}[!t]
\centering
\includegraphics{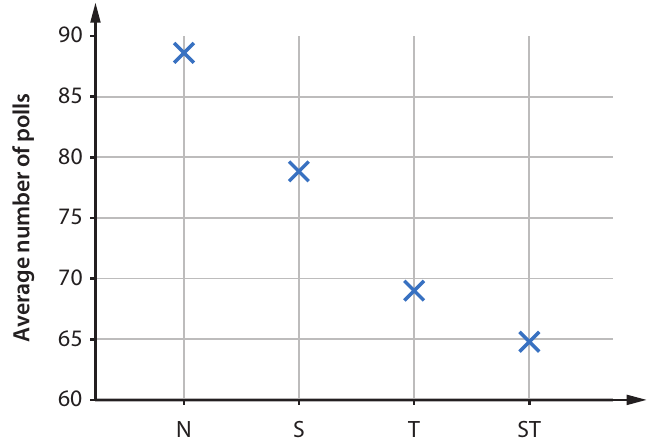}
\caption{Average number of polling messages sent by a \ac{UE}.}
\label{fig:polls}
\end{figure}

First, we measured the average number of polling messages sent by each \ac{UE} as shown in Fig.\;\ref{fig:polls}.
Since the 99\,\%~confidence intervals were narrow, we omitted them here.
From this figure, we can see that without a hierarchy (N), a \ac{UE} usually sends about 88~polling messages.
In contrast, when employing a hierarchy, we can see the expected decrease of polling messages.
For spatial aggregation (S), the number is reduced to about 79~messages.
We also see that temporal aggregation (T) is more effective than (S) with about 69~messages.
On one hand, this behavior can be explained by the fact that \acp{UE} tend to stay within the same cell for a longer time.
On the other hand, for (S), cells are only clustered randomly without considering the road network.
This may lead to cells being grouped into one cluster, although no \acp{UE} may travel between these cells.
Finally, with both spatial and temporal aggregation (ST), we can see a further decrease to about 65~messages.

\begin{figure}[!t]
\centering
\includegraphics{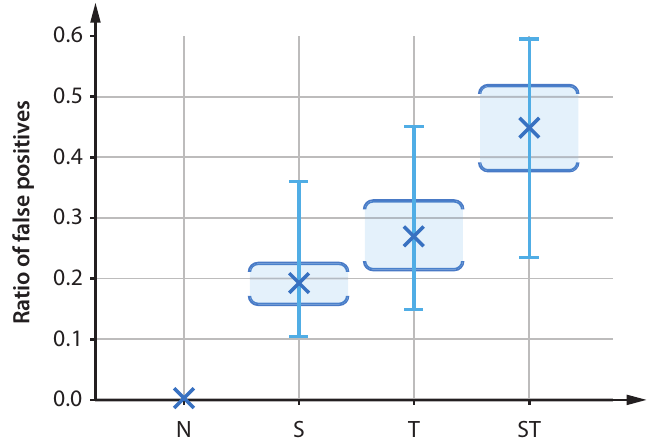}
\caption{Ratio of false positives among all messages received by \acp{UE}.}
\label{fig:falsepos}
\end{figure}

While the token hierarchy is able to provide the expected decrease of polling messages, it should also result in \acp{st-d} being received by \acp{UE} not having resided in the intended region.
Therefore, we measured both the total number of messages being received and the total number of false positives for all \acp{UE}.
We then divided the number of false positives by the total number of messages to obtain the ratio of false positives.
Fig.\;\ref{fig:falsepos} shows the false positives ratio with 99\,\% confidence intervals and min-max-whiskers.
According to our expectations, for (N), there are no false positives.
For (S), the false positives increase to about 20\,\% while (T) results in a ratio of about 27\,\%.
Furthermore, for (ST), we obtain the highest ratio of about 45\,\%.
This shows the tradeoff between the polling overhead and the delivery accuracy and corresponds to our observation that random spatial aggregation is less efficient than temporal aggregation.
Hence, the ratio of false positives is slightly higher for (T) than for (S).

Our evaluation shows the potential benefit of using a token hierarchy in order to reduce the polling overhead.
Therefore, we consider finding the optimal tradeoff between the polling overhead and the false positives ratio part of our future work.

\section{Conclusion}
\label{sec:conclusion}

In this article, we introduced the concept of \ac{STM}.
We provided an extensive overview of the envisioned applications and objectives to be fulfilled.
Moreover, we proposed a first approach to realize such a service and provided a detailed discussion of its privacy features.
Finally, we evaluated our approach in a large-scale simulation setup.
We showed the potential benefit of employing a token hierarchy which provides a tradeoff between delivery accuracy and polling overhead.

Several issues remain to be improved in further studies.
First of all, we plan to evaluate the performance of our approach regarding the functional and non-functional objectives in more detail.
Furthermore, instead of random clustering, optimization-based clustering strategies should be considered.
Finally, we plan to investigate the potential of distributed approaches for realizing an \ac{STM} service.

\section*{Acknowledgment}
This work is supported by the German Research Foundation (DFG Graduiertenkolleg~1487, Selbstorganisierende Mobilkommunikationssysteme f\"ur Katastrophenszenarien).

\bibliographystyle{IEEEtran}
\bibliography{IEEEabrv,references}

\end{document}